\documentstyle[twocolumn,floats,aps,prb,psfig]{revtex}

\ifx\undefined\psfig\def\psfig#1{ }\else\pssilent\fi
\begin{document}
\ifpreprintsty\else
\twocolumn[\hsize\textwidth%
\columnwidth\hsize\csname@twocolumnfalse\endcsname
\fi
\draft
\preprint{ }
\title {Dynamic exchange-correlation potentials for the electron gas
in \\ dimensionality $D=3$ and $D=2$} 
\author {R. Nifos\`\i, S. Conti,\cite{presentaddress} and M. P. Tosi}
\address {Istituto Nazionale di Fisica della Materia and Classe di
Scienze, Scuola Normale Superiore, I-56126 Pisa, Italy}
\date{\today} 
\maketitle

\begin{abstract}
Recent progress in the formulation of a fully dynamical local
approximation to time-dependent Density Functional Theory appeals to
the longitudinal and transverse components of the exchange and
correlation kernel in the linear current-density response of the
homogeneous fluid at long wavelength. Both components are evaluated
for the electron gas in dimensionality $D=3$ and $D=2$ by an
approximate decoupling in the equation of motion for the current
density, which accounts for processes of excitation of two
electron-hole pairs. Each 
pair is treated in the random phase approximation, but the role of
exchange and correlation is also examined; in addition, final-state
exchange processes are included phenomenologically so as to satisfy
the exactly known high-frequency behaviours of the kernel. 
The transverse and longitudinal spectra involve 
the same decay channels and are similar in shape.
A two-plasmon threshold in the spectrum for two-pair excitations in
$D=3$ leads to a sharp minimum in the real part of the exchange and
correlation kernel at twice the plasma frequency. In $D=2$ the same
mechanism leads to a broad spectral peak and to a broad minimum in the
real part of the kernel, as a consequence of the dispersion law of the
plasmon vanishing at long wavelength. The numerical 
results have been fitted to simple analytic functions.
\end{abstract}

\ifpreprintsty\else\vskip1pc]\fi
\narrowtext

\section{Introduction}
The plasmon dispersion relation and the dynamic structure factor of
the conduction electrons in simple metals are known from Electron
Energy Loss and Inelastic X-ray Scattering
experiments.\cite{Fink,Larsen96} Both electron-gas correlations and
band structure enter in determining these properties and
Time-Dependent (TD) Density Functional Theory (DFT) provides a general
framework which can account for both. The generality of the DFT
method\cite{RungeGross,Gross96} in dealing with inhomogeneous electron
systems motivates interest in the derivation of sensible
approximations to the dynamic exchange and correlation (xc) potential,
beyond the adiabatic
regime\cite{Zangwill80,Zangwill81,Ekardt84,Ekardt85} whose
applicability is limited to low-frequency phenomena. 

A local density approximation for a scalar xc potential in
time-dependent phenomena was proposed in early work of Gross and
Kohn.\cite{GrossKohn85,IGK} However, detailed analysis of the
constraints coming from basic conservation laws has shown that
inconsistencies can arise and are associated with the non-existence of
a gradient expansion for the frequency-dependent xc potential in terms
of the density
alone.\cite{Gross96,Dobson94,Dobson94b,Vignale95b,Vignale95a} Recently
Vignale and Kohn\cite{Vignale96,VignaleKohn96b} have overcome these
difficulties by resorting to a dynamic xc {\em vector} potential even
in the case of an inhomogeneous system subject to an external {\em scalar}
potential. They obtained an explicit local-density expression for the
xc vector potential in the linear response regime in terms of
correlations of longitudinal and transverse currents in the
homogeneous electron gas. This expression becomes exact when the
equilibrium electron density and the external potential are slowly
varying in space, on length scales set by $k_F^{-1}$ and $v_F/\omega$
where $k_F$ and $v_F$ are the local Fermi wave-number and
velocity. 

The results of Vignale and Kohn have been interpreted
in terms of complex, frequency-dependent viscoelastic coefficients,
allowing a non-linear generalization up to second order in the spatial
gradients.\cite{VUC97} Within this framework, Ullrich and
Vignale\cite{UV98} have developed a simple computational 
scheme for the  calculation of the linewidths of non-Landau-damped 
collective excitations, whose decay is solely due to dynamical xc effects.
The order-of-magnitude
enhancement over the homogeneous-gas approximation reported in the
application to quantum strips\cite{UV98} emphasizes the relevance of
the interplay between inhomogeneity and dynamical xc effects.

The work of Vignale and Kohn has brought new interest to the
evaluation of the longitudinal and transverse components of the
dynamic xc kernel [${f_{xc}^L}(\omega)$ and ${f_{xc}^T}(\omega)$, say] in the
homogeneous electron gas at long wavelength. An interpolation between
the known asymptotic behaviours at low and high frequency was proposed
for the longitudinal component by Gross and Kohn\cite{GrossKohn85,IGK}
and later extended to finite wave number by
Dabrowski.\cite{Dabrowski86} It was subsequently noticed\cite{BCT96}
that the long-wavelength longitudinal spectrum is closely related to
processes of excitation of two correlated electron-hole pairs in the
electron gas. These have been studied by perturbative methods in the
early work of Dubois\cite{Boi59,BoK69} and by several other
authors.\cite{NPS66,HW,GlickLong,Gas84,Gas92,BMM91,BBS93b} The
inclusion of dynamic screening in the Random Phase Approximation (RPA)
leads to a mode-coupling form of the spectrum\cite{HW} and inclusion
of final-state exchange processes is needed to recover the exact
high-frequency behaviour calculated by Glick and
Long.\cite{GlickLong} A similar expression has been more recently
obtained by Neilson {\em et al.}\cite{Neilson} within a memory
function formalism for the electron gas in dimensionality $D=2$. 

With regard to the transverse component of the long-wavelength xc
kernel ${f_{xc}^T}(\omega)$, Vignale and Kohn\cite{Vignale96,VignaleKohn96b}
have given a first-order perturbative estimate and obtained the
high-frequency limit. We have subsequently briefly reported on the
results of a full calculation of ${f_{xc}^T}(\omega)$ for the electron gas in
three spatial dimensions,\cite{CNT97} and on preliminary results
in the two-dimensional case.\cite{NCT97}
These calculations were based on the two-pair model treated in the
RPA and corrected for final-state exchange processes (hereafter
indicated as RPAE). 

In the present work we give a full account of our approach to the
evaluation of the dynamic xc kernels ${f_{xc}^L}(\omega)$ 
and ${f_{xc}^T}(\omega)$,
including an exact expression for ${f_{xc}^{L,T}}(\omega)$ 
in terms of four-point
response functions. We also extend our calculations to 
(i) fully evaluate the dynamic xc kernels for the electron gas with $e^2/r$
interactions in $D=2$, and (ii) examine the role of including
exchange and correlation in the screening processes entering the RPAE.

The lay-out of the paper is as follows. Section \ref{secanres}
presents the theory 
underlying our calculations, with the help of three Appendices. We start
from the definition of ${f_{xc}^{L,T}}(\omega)$ in terms of the current-current
response functions for the ideal and the real electron gas and proceed
to evaluate the ideal-gas response at high frequency and to derive an
exact expression
for the real-gas response in terms of a four-point response
function. An approximate decoupling of this latter function into
products of two-point response functions introduces the two-pair
approximation. Screening at the RPA level or better and
phenomenological inclusion of final-state exchange finally lead to
the formulae used in our calculations.  The numerical results are
presented in Section \ref{secrisRPA} together with fits to analytic functions
incorporating the known asymptotic behaviours and aimed at
facilitating numerical applications within the time-dependent DFT
formalism. The role of exchange and correlation in the treatment of
each pair is studied in Section \ref{secrisSTLS}. We conclude with a
brief summary in Section \ref{secconcl}. 

\section{Theory: exact results and two-pair model}
\label{secanres}
The longitudinal (L) and transverse (T) kernels ${f_{xc}^{L,T}}(\omega)$ of the
homogeneous electron gas are defined as the $k\to0$ limit of the
functions
\begin{eqnarray}
f_{xc}^{L,T}(k,\omega) &=& {\omega^2\over k^2} \left[
 {1\over \chi^0_{L,T}(k,\omega)+n/m}\right.\nonumber\\
&&-\left.{1\over\chi_{L,T}(k,\omega)+n/m} \right] -v_k^{L,T}\,.
\label{eqdeffxltchi}
\end{eqnarray}
Here, $\chi_{L(T)}$ is the longitudinal (transverse)  current-current
response function of the homogeneous fluid at density $n$,
$\chi^0_{L(T)}$ is the corresponding ideal-gas response function,
$v_k^L$ is the Coulomb potential ($v_k^L = 4\pi e^2/k^2$ in $D=3$ and
$v_k^L=2\pi e^2/k$ in $D=2$) and $v_k^T=0$. 
We remark \cite{PinesNoz,SingwiTosi} that $\chi_L(k,\omega)$ is related to
the density-density response function $\chi(k,\omega)$ by
\begin{equation}
\label{eqchilchirho}
\chi_L(k,\omega) + {n\over m} = {\omega^2\over k^2} \chi(k,\omega)
\end{equation} 
and that ${f_{xc}^L}$ is proportional to the local field factor $G(k,\omega)$
entering the dielectric function of the electron gas, according to
${f_{xc}^L}(k,\omega) = - v_k^L G(k,\omega)$. 

The longitudinal response function $\chi^0_L(k,\omega)$ is immediately
obtained from the Lindhard
susceptibility\cite{PinesNoz,SingwiTosi,Lindhard54} in 
$D=3$ and from the Stern susceptibility\cite{Stern} in $D=2$ by using
(\ref{eqchilchirho}) for the ideal Fermi gas. The calculation of
$\chi^0_T(k,\omega)$ for the ideal Fermi gas in $D=3$ and $D=2$ is
reported in Appendix \ref{applindhT}. 

Equation (\ref{eqdeffxltchi}) may be written in terms of the proper
current response functions $\tilde\chi_{L,T}(k,\omega)$,
\begin{eqnarray}
f_{xc}^{L,T}(k,\omega) &=& {\omega^2\over k^2} \left[
 {1\over \chi^0_{L,T}(k,\omega)+n/m}\right. \nonumber\\
&&\hskip5mm\left.-{1\over\tilde\chi_{L,T}(k,\omega)+n/m} \right] \,.
\label{eqdeffxltchiprop}
\end{eqnarray}
This emphasizes that the plasmon does not contribute to ${f_{xc}^L}(k,\omega)$,
leaving only contributions from multi-pair excitations in the limit
$k\to0$. In fact, to leading order in the long wavelength limit we
have from (\ref{eqdeffxltchiprop}) 
\begin{equation}
\label{eqfxcimchitilde}
{\bf Im} {f_{xc}^{L,T}}(\omega) = \lim_{k\to0} 
{m^2\omega^2\over n^2 k^2} {\bf Im}
\tilde\chi_{L,T}(k,\omega) \,.
\end{equation}
We proceed below to evaluate the imaginary part of the kernels, from
which their real part will be obtained by means of the Kramers-Kronig
relation.
The equation of motion for $\chi_{ij}({\bf k},\omega)$ can be obtained
from the definition of a general response function in terms of
unequal-time commutators,  
\begin{equation}
{\left\langle\!\left\langle {A}; {B} \right\rangle\!\right\rangle_{\omega}} 
\,=-i\int_{0}^{\infty}
e^{i(\omega+i\epsilon)t}\langle[A(t),B(0)]\rangle dt
\end{equation}
where $A(t)=e^{iHt}Ae^{-iHt}$, $\langle\dots\rangle$ denotes a 
ground-state expectation value, and $\epsilon$ is a positive
infinitesimal. The current-current response  
$\chi_{ij}({\bf k},\omega)=\,\langle\langle {\bf j}_{{\bf k}}^{i};
{\bf j}_{-{\bf k}}^{j}\rangle\rangle_{\omega}$
satisfies the equation of motion
\begin{equation}
\omega^2\chi_{ij}({\bf k},\omega)= \left\langle\left[[{\bf j}_{\bf k}^i,
H],{\bf j}_{\bf -k}^j\right]\right\rangle 
-{\left\langle\!\left\langle {[{\bf j}_{\bf k}^i,H]};  
{[{\bf j}_{\bf -k}^j,H]} \right\rangle\!\right\rangle_{\omega}} 
\label{eqeqmotchiij}
\end{equation}
where the first term is real and independent of $\omega$. It  will be
evaluated below in the discussion of the real part of $f_{xc}$ (Eqs.\
\ref{M3Leq}-\ref{eqfxclwinfty}).  
Quite lengthy calculations, which are briefly reported in Appendix
\ref{appcommsmallk}, lead to the long-wavelength result
\begin{eqnarray}
\label{eqesatta}
{\bf Im}\tilde\chi_{ij}({\bf k},\omega)&=&
\frac{1}{m^{2}V^{2}\omega^{4}}\sum_{{\bf q},{\bf q}'}
{\bf Im}\left\langle\!\left\langle {{\bf j}_{{\bf q}}^{l}\rho_{-{\bf
          q}}}; 
{{\bf j}_{{\bf q}'}^{l'}\rho_{-{\bf q'}}} 
\right\rangle\!\right\rangle_{\omega}\nonumber\\
&\times& \Gamma^{il}({\bf q},{\bf k})\Gamma^{jl'}({\bf q'},-{\bf k})+o(k^{2})
\end{eqnarray}
where the coefficients $\Gamma^{il}({\bf q},{\bf k})$ are defined in
Eq.~(\ref{eqdefGamma}) and summation over repeated indices is
understood.  This expression, together with
(\ref{eqfxcimchitilde}), gives an exact result for ${\bf Im}
f_{xc}^{L,T}(\omega)$. 

We now discuss some approximate evaluations of Eq.~(\ref{eqesatta}). 
Use of the ideal-gas four-point response function in the RHS 
gives the exact second-order perturbative value
for ${f_{xc}^{L,T}}(\omega)$, which is expected to be accurate at high
frequency (see App. \ref{apppert}).  
In Appendix \ref{apppert} we derive in this way the leading high-frequency
behaviour of the kernel,
\begin{equation}\label{eqhighw}
{f_{xc}^{L,T}}(\omega) = - a_{L,T} \pi^2 \left(2\over\pi\right)^{D-2} \left(2
Ry\over \omega\right)^{D/2} a_B^D Ry\end{equation}
where $a_B$ is the Bohr radius, $a_L=23/30$ and $a_T=8/15$ in $D=3$, while
$a_L=11/16$ and $a_T=9/16$ in $D=2$. The longitudinal component
of this result was obtained by Glick and Long\cite{GlickLong} in
$D=3$ and by Holas and Singwi\cite{Holas89} in $D=2$. 
In the low-frequency limit instead, perturbation theory gives an
artificial divergence in ${\bf Re} f_{xc}$ and a discontinuity in
${\bf Im } f_{xc}$, and will not be pursued further.

We obtain an approximate  
nonperturbative evaluation of (\ref{eqesatta}) by decoupling 
its RHS within an RPA-like scheme, 
which in the frequency domain gives
\begin{eqnarray}
\label{decoup}
{\bf Im}\left\langle\!\left\langle {AB}; {CD}
  \right\rangle\!\right\rangle_{\omega}
&\simeq&-\int_0^\omega
 \frac{d\omega'}{\pi}\left[{\bf Im}
\left\langle\!\left\langle {A}; {C}
  \right\rangle\!\right\rangle_{\omega'}\right.
\label{eqdecouplingabcd}\\
&&\hskip-2.8cm\left.\times{\bf Im}
\left\langle\!\left\langle {B}; {D} 
\right\rangle\!\right\rangle_{\omega-\omega'}
 +{\bf Im}
\left\langle\!\left\langle {A}; {D} \right\rangle\!\right\rangle_{\omega'}
{\bf Im}
\left\langle\!\left\langle {B}; {C}
  \right\rangle\!\right\rangle_{\omega-\omega'}\right]\,. 
\nonumber
\end{eqnarray}
The functions in the RHS of Eq.\ (\ref{decoup}) are understood to be exact
response functions of the interacting electron gas. 
This scheme clearly neglects
exchange processes in the final state, 
which are known to reduce the spectral strength by
a factor of 2 at high frequency in perturbative treatments
(see Appendix \ref{apppert}), but are ineffective at low frequency.
This can be physically understood as follows. A two-pair
excitation at long wavelength involves the creation of holes with
momenta ${\bf p}$ and ${\bf p'}$ inside the Fermi sphere and of
electrons with momenta  ${\bf p}+{\bf q}$ and ${\bf p'}-{\bf q}$
outside the Fermi sphere. 
If the excitation energy $\omega$ is small compared to the
Fermi energy $\varepsilon_F$, then necessarily $|{\bf q}|\ll k_F$ and,
since on average
$|{\bf p}-{\bf p'}|\simeq k_F$, each electron is substantially closer (in
${\bf k}$-space) to ``its'' hole than to the other one: exchange
processes are thus suppressed in this case (see Fig.~\ref{figfs}.a) by
a factor $v^L_{k_F}/v^L_q$ which vanishes for $q\to 0$ (i.e. $\omega\to
0$).\cite{footnotescreendpot}
Conversely if
$\omega\gg \varepsilon_{F}$ 
one has $|{\bf q}|\gg k_F$ and therefore for parallel 
spins the strengths of direct and exchange processes are equal and
opposite  (see Fig.~\ref{figfs}.b). 

On the basis of the above argument, which can be made 
quantitative at a perturbative level, we phenomenologically
incorporate exchange effects by inclusion of a factor
\begin{equation}
g_x(\omega)={1  + 0.5 \omega/2\varepsilon_F
\over 1+ \omega/2\varepsilon_F}\,,
\label{eqdefgx}
 \end{equation}
which interpolates between $1$ at low $\omega$ and $0.5$ at high $\omega$ on
the energy scale ($2 \varepsilon_F$) of exchange processes.
Our final expression for the imaginary part of the kernel thus is
\begin{eqnarray}
{{\bf Im}}f_{xc}^{L,T}(\omega) &=&
- g_x(\omega) \int^\omega_0{d\omega'\over\pi}
\int{d^Dq\over(2\pi)^D n^2}  (v_q^L)^2 \nonumber\\
&&\hskip-1.5cm\times
\left[a_{L,T}{q^2\over\omega'^2}{{\bf Im}}\chi_L(q,\omega')+
b_{L,T}{q^2\over\omega^2} {{\bf Im}}\chi_T(q,\omega')\right]
\nonumber\\
&&\hskip-1.5cm\times {q^2\over(\omega-\omega')^2} {{\bf Im}}
\chi_L(q,\omega-\omega'),
\label{eqimchil}
\end{eqnarray}
with $a_{L,T}$ as defined below Eq.~(\ref{eqhighw}), $b_L=8/15$ and
$b_T=2/5$ in  $D=3$ and  
$b_L=b_T=1/2$ in $D=2$. 
The expression for the longitudinal component is equivalent to that
obtained in 3D  by Hasegawa and Watabe \cite{HW} by diagrammatic
means and  
similar to that derived in  2D by Neilson \textit{et al.} \cite{Neilson}
(see the discussion in Ref. \onlinecite{BCT96}). 
This result can be physically understood as representing the spectral
density of two-pair excitations, which are present at any
frequency in the long-wavelength region of the spectrum and provide a
mechanism for the decay of the plasmon outside of the single-pair
continuum. The resulting linewidth is   $\Gamma_k/\omega_k
= - {\bf Im} f_{xc}^L(\omega_k) / v_k^L$. 

Considering the low-fre\-quen\-cy behaviour of ${\bf Im}\,
\chi_{L,T}(k,\omega)$ it can be shown that at low $\omega$
Eq.~(\ref{eqimchil}) is linear in $\omega$, and that to this order
the first term in the square brackets does not contribute. 
The coefficients of this linear behaviour are related\cite{VUC97} to the 
bulk and shear viscosities of the electron gas, $\zeta$ and $\eta$
respectively, via 
\begin{equation}
  \zeta = - n^2 \lim_{\omega\to0} \left[ {{\bf Im}
  f_{xc}^L(\omega)\over \omega} - 2 {D-1\over D} {{\bf Im}
  f_{xc}^T(\omega) \over\omega}\right] \label{eqdefzeta}  
\end{equation}
and
\begin{equation}
  \eta = - n^2 \lim_{\omega\to0} {{\bf Im}
  f_{xc}^T(\omega)\over \omega} \label{eqdefeta} \,.
\end{equation}
The significance of $\zeta$ and $\eta$ is the same as in
classical hydrodynamics.\cite{Landau6} Both viscosities vanish in the
ideal Fermi gas, as well as in the RPA and in static-local-field
approximations to the interacting electron gas. 
Since $b_L/b_T = 2 (D-1)/D$, also within the present model the bulk
viscosity $\zeta$ vanishes identically. Numerical results for the shear
viscosity $\eta$ will be given below.  

We now discuss the real part of the kernels
$f_{xc}^{L,T}(\omega)$. Once the imaginary part has been evaluated, the
real one can be obtained by means of the Kramers-Kronig relation
\begin{equation}
{\bf{Re}} {f_{xc}^{L,T}}(\omega)={f_{xc}^{L,T}}(\infty)+
\frac{1}{\pi} \int_{-\infty}^{+\infty}d\omega'
\frac{{{\bf Im}}{f_{xc}^{L,T}}(\omega')}
{\omega'-\omega}\,,
\label{KK1}
\end{equation}
where ${f_{xc}^{L,T}}(\infty)$ denotes the (real) high-frequency limit
of the kernels. This quantity corresponds to the long-wavelength value of  the
leading spectral moment beyond the $f$-sum rule,\cite{PinesNoz,SingwiTosi} 
which is the first term on the RHS of Eq.~(\ref{eqeqmotchiij}).
With the definition $\tilde M_{L,T}(k)=\lim_{\omega\to\infty}
\omega^2\tilde\chi_{L,T}(k,\omega)$, we have
\begin{eqnarray}
\tilde{M}_{L}(k)&=&\frac{n
k^{2}}{2m^{2}}\left[\frac{k^{2}}{2m}+
\frac{12}{D}{\langle ke \rangle}\right.\nonumber\\
&&\hskip-0.7cm\left.+\frac{2}{n}\sum_{{\bf q}} v_k^{L}
\frac{({\bf k}\cdot{\bf q})^{2}}
{k^{4}}\left(S(|{\bf q}+{\bf k}|)-S(q)\right)\right]\,,
\label{M3Leq}
\end{eqnarray}
and
\begin{eqnarray}
\tilde{M}_{T}(k)&=&\frac{n
k^{2}}{2m^{2}}\left[\frac{4}{D}{\langle ke \rangle} \right.\nonumber\\
&&\hskip-1.2cm\left.+\frac{1}{n}\sum_{{\bf q}} v_k^{L}
\left[\frac{q^{2}}{k^{2}}-\frac{({\bf k}\cdot{\bf q})^{2}}
{k^{4}}\right]
\left(S(|{\bf q}+{\bf k}|)-S(q)\right)\right]
\label{M3Teq}
\end{eqnarray}
where $S(q)$ is the static structure factor and ${\langle ke \rangle}$
denotes the true kinetic energy per particle. 
Expansion at long wavelengths gives the required high-$\omega$ limit of
${f_{xc}^{L,T}}(\omega)$, 
\begin{eqnarray}
f_{xc}^{L,T}(\infty) = 
\frac{1}{2n}\left[d_{L,T}
({\langle ke \rangle}-{\langle ke \rangle}^0)
+e_{L,T}{\langle pe \rangle}\right]
\label{eqfxclwinfty}
\end{eqnarray}
with $d_L=4$, $d_T=4/3$, $e_L=8/15$ and $e_T=-4/15$ in $D=3$, and
$d_L=6$, $d_T=2$, $e_L=5/4$ and $e_T=-1/4$ in $D=2$.
The average kinetic and potential energies 
${\langle ke \rangle}$ and ${\langle pe \rangle}$ can be
obtained from the Monte Carlo equation of
state,\cite{CeperleyAlder,VWN80,Rapisarda96}
and ${\langle ke \rangle}^0$ denotes the ideal-gas value. The
resulting values of $f_{xc}^{L,T}(\infty)$, given in Table
\ref{tablefxcasy} ($D=3$) and Table \ref{tabvisco2d} ($D=2$), allow to
evaluate numerically Eq.~(\ref{KK1}) and to obtain the real part of
the kernels at any frequency. 

The low-frequency limit of $f_{xc}^{L,T}(\omega)$ is related\cite{VUC97} to
the elastic moduli $K$ and $\mu$ via
\begin{equation}
  K_{xc} = n^2\left[ f_{xc}^L(0) - 2 {D-1\over D}
  f_{xc}^T(0)\right] \label{eqdefK}
\end{equation}
and
\begin{equation}
  \mu_{xc} = n^2 f_{xc}^T(0) \label{eqdefmu} \,.  
\end{equation}
The significance of $K$ and $\mu$ is the same as in
classical elasticity,\cite{Landau7}
and as usual the suffix $xc$ indicates that the ideal-gas contribution
has been subtracted.  Our results  for $K_{xc}$ are in good
agreement with the accurate values obtained from the Monte Carlo $xc$
energy per 
particle\cite{CeperleyAlder,VWN80,Rapisarda96} 
$\epsilon_{xc}^{\mathrm{MC}}(n)$ via $K_{xc}^{\mathrm{MC}} = n^2 d^2 
\epsilon_{xc}^{\mathrm{MC}}/ dn^2$ (see below\cite{fxclungobetabar}).
We finally note that $K_{xc}$ is also related to the
long-wavelength limit of the {\em static} $f_{xc}^L(k,0)$ via the
compressibility sum rule,\cite{PinesNoz}
\begin{equation}
 K_{xc} = n^2 \lim_{k\to0}\lim_{\omega\to0}f^{L}_{xc}(k,\omega)
 \,. \label{eqfxclw0} 
\end{equation}
  
\section{Numerical results within the RPA}
\label{secrisRPA}
This Section presents results that we have obtained for  $f_{xc}$ from
numerical integration of Eqs.~(\ref{eqimchil}) and (\ref{KK1})  
using RPA response
functions, which are given by
\begin{eqnarray}
 {1\over \chi^{RPA}_{L,T}(k,\omega)+n/m}=
{1\over\chi^{0}_{L,T}(k,\omega)+n/m} -{k^2\over \omega^2} v_k^{L,T}\,.
\label{eqchiRPA}
\end{eqnarray}
In the high-frequency limit the first
term in the square brackets of Eq.(\ref{eqimchil}) dominates over the second
one, implying  ${\bf Im} f_{xc}^T = (a_T/a_L) {\bf Im} f_{xc}^L$;
conversely, for low $\omega$ the first term is negligible
and ${\bf Im} f_{xc}^T = (b_T/b_L){\bf Im} f_{xc}^L$.
Indeed the longitudinal and transverse spectra are very similar, and
the transverse one is rather accurately reproduced at all
frequencies by setting ${\bf Im} f_{xc}^T(\omega) \simeq 0.72 \cdot
{\bf Im} f_{xc}^L(\omega)$ in 3D and  
${\bf Im} f_{xc}^T(\omega) \simeq 0.85 \cdot
{\bf Im} f_{xc}^L(\omega)$ in 2D, the proportionality factor being 
close both to $a_T/a_L$ and to $b_T/b_L$. 
For the real part of the kernels there is an additional shift due to 
the different values for $\omega=\infty$. 

Figure \ref{figim3} reports the results for
the imaginary part of ${f_{xc}^{L,T}}$ in 3D at $r_s=3$ 
as functions of $\omega/\omega_{pl}$ 
[in 3D $r_s$ is defined as $(4\pi n a_B^3/3)^{-1/3}$ and 
the plasma frequency  $\omega_{pl}$ as 
$(4\pi e^2 n/m)^{1/2}$]. 
Both our results and the Gross-Kohn (GK)
 inter\-po\-la\-tion\cite{GrossKohn85,IGK} 
 for ${\bf Im} {f_{xc}^{L,T}}$ reproduce the high
frequency behaviour given by 
Eq.~(\ref{eqhighw}) and are linear at low frequency. 
As already remarked the bulk viscosity $\zeta$ vanishes identically within
the present model, but the shear viscosity $\eta$ is finite; numerical
results for the latter are given in Table \ref{tabvisco3D}. 
{}From Fig.\ \ref{figim3} one can see that our estimate for $\eta$ is
 significantly smaller than that of GK. In the intermediate
frequency region
our curves exhibit a sharp threshold at twice the plasma frequency,
which is due to the onset of two-plasmon processes. This is the most
remarkable difference from the GK interpolation.

The corresponding real parts of the kernels are shown in Fig.~\ref{figre3}.
The two-plasmon threshold in the spectrum generates
a pronounced minimum at
$\omega=2{\omega_{pl}}$ in the real part, 
which is absent in the smooth GK interpolation. 
The $f_{xc}^L(0)$ value was obtained by GK assuming
$f_{xc}^L(0) = K_{xc}/n^2$, i.e.\ $f_{xc}^T(0)=\mu_{xc}=0$ [see
Eqs. (\ref{eqdefK}-\ref{eqdefmu})]; 
our curves instead are consistent with Eq.\ (\ref{eqdefK}) with a
 nonvanishing $f_{xc}^T(0)$.
Table \ref{tabvisco3D}  reports our results for the
elastic moduli $K_{xc}$ and $\mu_{xc}$, obtained from ${\bf Re}
f_{xc}$ via Eqs.~(\ref{eqdefK}-\ref{eqdefmu}). They cannot be expected to be
very precise, being obtained through integration over the entire spectrum.
With this {\em caveat}, we note that our 
numerical estimates for $K_{xc}$ agree with the accurate Monte Carlo
results $K_{xc}^{\mathrm{MC}}$, also given in the same Table, within $5\%$.
We note that the xc contribution to the shear modulus $\mu_{xc}$ becomes
negative at low 
density; the total shear modulus $\mu=\mu_{xc}+{2\over5} n
\varepsilon_F$, however, remains positive. 

In order to facilitate use of these data in practical TD-DFT
computations, an analytical fit has been given in 
Ref.~\onlinecite{CNT97}. The constraint $ f_{xc}^L(0)=K_{xc}/n^2$,
which was imposed in the fits following GK, can be removed simply by
 setting $\beta=1$.  

We now turn to the two-dimensional system. Figures \ref{fig2dim2} and
\ref{fig2dre2} report our results 
for the imaginary and real parts of ${f_{xc}^{L,T}}$ at $r_s=3$ [in 2D $r_s$ is
defined as $(\pi n a_B^2)^{-1/2}$]. 
The main difference with respect to the 3D results is the absence of the
sharp two-plasmon threshold in the two-pair excitation spectrum  (i.e.\
in ${\bf Im} f_{xc}$), due to the fact that the plasmon dispersion relation 
vanishes in 2D at long wavelength. Correspondingly the minimum
in the real part of the kernel is much broader. 
The figures also compare our result for ${f_{xc}^L}$ 
with the interpolation scheme of Holas and Singwi (HS),\cite{Holas89}
which is the 2D extension of the GK interpolation.
Both curves reproduce the asymptotic limit (\ref{eqfxclwinfty})
as well as the high-frequency behaviour of Eq.~(\ref{eqhighw}), and
both imaginary 
parts are linear in $\omega$ at low frequency.  
Also in the 2D case the minimum in the real part is absent in
the GK/HS interpolation scheme.

Table \ref{tabvisco2d} reports the resulting values of the shear
viscosity $\eta$ and of the elastic moduli $K_{xc}$ and $\mu_{xc}$,
obtained as in 
the 3D case. In 2D the agreement between the bulk modulus $K_{xc}$ obtained
from our data on $f_{xc}^{L,T}$ and the Monte Carlo value
$K^{\mathrm{MC}}_{xc}$ is 
not as good as in 3D, but still better than $10\%$ at all values of  $r_s$ 
that we have considered.
In contrast to 3D, at low density the total shear modulus
$\mu=\mu_{xc}+{1\over2} n \varepsilon_F$, turns out to be
negative. However, the observed 
disagreement between $K_{xc}$ and $K_{xc}^{MC}$ prevents us from
drawing a conclusion about the presence of an instability.

To facilitate application of these results in actual TD-DFT
computations, such as the one of Ref.~\onlinecite{UV98}, we provide
below a simple analytical interpolation. The 
imaginary part is reproduced by
\begin{equation}\label{eqfxcfit2d}
{\bf Im} {f _{xc}^L}(\omega) = - g_x(\omega)
{c_1 \omega+ c_2 \omega^2+c_3 \omega^3 +2 c_{HS} \omega^5
  \over c_0+c_4  \omega^4+\omega^6} 
\end{equation}
where $\omega$ is in $Ry$, $f_{xc}$ in units of $Ry/n$ and 
$c_{HS}=11\pi/8r_s^2$ is the coefficient of the leading high-frequency
behaviour from Eq.~(\ref{eqhighw}). The remaining parameters, which we
obtained by a least-squares fit to the numerical calculation, are
reported in Table \ref{tabparfit2d}; the transverse component can be
approximated by ${\bf Im}{f_{xc}^T}=0.85\,{\bf Im}{f_{xc}^L}$. The
quality of a typical fit is shown in Figures \ref{fig2dim2} and
\ref{fig2dre2}. 
Values at intermediate $r_s$ are best obtained by interpolation of
$f_{xc}(\omega)$ at the same $\omega$ (in $Ry$); for $r_s\ge 3$ also
the simpler 
scheme of interpolating the fitted coefficients $c_0,\dots,c_4$ is viable.
The real part can be obtained from Eq.~(\ref{KK1}), which in this case
can be integrated analytically; the resulting expression is quite long
and we do not report it. 

\newbox\tmp
\setbox\tmp=\hbox{{$f_{xc}$}}
\section{Exchange-correlation effects\\
on \box\tmp}
\label{secrisSTLS}
This Section is aimed at
assessing the validity of the results presented above.
In the first part we  
study the effect of correlation in the treatment
of each pair, adopting more refined response functions in the RHS of
Eq.~(\ref{eqimchil}); in the final part we discuss  
corrections which go beyond the present two-pair
 model.

We have introduced the effect of correlation in the treatment of each
pair in Eq.~(\ref{eqimchil}) by means of two of the most
successful static-local-field approximations, the
Singwi-Tosi-Land-Sj\"olander\cite{STLS} (STLS) and the
Vashishta-Singwi\cite{VS} (VS). In the 3D case, both schemes predict
negative plasmon dispersion at large $r_s$ ($r_s\ge5$ in STLS and 
$r_s\ge 9$ in VS), in qualitative agreement
with experiment.\cite{Fink} The VS scheme embodies the compressibility sum rule
on the static response, and is therefore more reliable in the study of
static phenomena. Since these schemes only involve longitudinal
currents, the transverse response function is still evaluated at an
RPA level. 

Figures \ref{figconfim15} and \ref{figconfre15} compare the 
RPA results in 3D with those obtained
with STLS and VS. At $r_s=1$
correlation only gives minor corrections, but at larger $r_s$ it
leads to a divergence, caused by the appearance of negative plasmon
dispersion. Figure \ref{figstlsrs} compares the results obtained
with STLS at various densities. The minimum at
intermediate frequency gets more pronounced with increasing $r_s$, and
for $r_s\ge5$ becomes  a divergence of the form
$\theta(\omega-2\omega_{min})(\omega-2\omega_{min})^{-1/2}$, where
$\omega_{min}$ is the minimum energy of the collective mode.

Introduction of the local field correction significantly increases the
shear viscosity $\eta$, and leads to large negative values of the shear
modulus $\mu_{xc}$ (e.g. with VS we get $\mu_{xc} = -0.03$ at $r_s=5$ and
$\mu_{xc}=-0.1$ at $r_s=10$, in units of $2 \omega_{pl} n$). The
relation $K_{xc} = K^{\mathrm{MC}}_{xc}$ is satisfied with the same
accuracy as in the RPA case.

Figure \ref{figskwstlsrpa} displays the dynamical structure factor
$S(k,\omega) = -(2k^2/n\omega^2) {\bf Im} \chi_L(k,\omega)$ which is
relevant to 
inelastic scattering experiments. The threshold behaviour at frequency
$2{\omega_{pl}}$ is a clear-cut signature of the present results for $f_{xc}$.
The more pronounced singularity present in the STLS curves originates
from the negative plasmon dispersion.

We also investigated the role of exchange and correlation in 2D, using
the STLS model as generalized in 2D by Jonson.\cite{Jonson76} The
qualitative behaviour turns out to be similar. Figure
\ref{fig_rpastls2dim} reports our results for the imaginary part of
$f_{xc}$ as obtained from RPA and STLS calculations. As in 3D, in
STLS the plasmon energy at intermediate wavevector is lower than in
RPA, and correspondingly we note a significant increase in the depth
of the minimum at intermediate frequency in ${\bf Im} f_{xc}$. This
increase, unlike the 3D case, is significant also at rather high
density (see the inset) as a consequence of
the enhancement of correlation effects in 2D systems. On the other
hand, in 2D no divergence appears in the two-pair spectrum, since no minimum at
finite $k$ is present in the plasmon dispersion.
The same is true for the corresponding real parts, which are shown in
Fig.\  \ref{fig_rpastls2dre}. 

Having discussed at length the role of two-pair processes in the
long-wavelength spectrum of the uniform electron gas, we now turn to a
qualitative discussion of other effects which have been neglected
in the present treatment.
Whereas a detailed quantitative calculation is cumbersome, the
qualitative role of multi-pair processes beyond two-pairs can be
easily grasped by straightforward extension of the present treatment,
in the spirit of an expansion in the number of pairs involved.
A perturbative analysis shows that in the high-frequency limit these
processes are of higher order in $1/\omega$, and therefore do not
contribute to the asymptotic behaviours of Eq.\ (\ref{eqhighw}). In the
intermediate-frequency regime one can expect the appearance of a
$n$-plasmon threshold effect in the $n$-pair channel. Whereas at
present we do not have a reliable quantitative estimate of the
contribution of such processes to ${\bf Im} f_{xc}$, we believe that
their overall 
spectral strength will be a minor correction to the present results.
This can be inferred from the good agreement between the two-pair
result for $K_{xc}$ and that from the Monte Carlo data, the
difference being a quantitative measure of the 
integrated spectral strength of higher-order processes. 

A different class of effects which could modify the present results
are improved response functions in the two-pair channel, i.e. in
Eq.\ (\ref{eqimchil}). 
The main qualitative feature which is absent from all response
functions that we have considered is plasmon damping. 
This will broaden the sharp feature
present in 3D at the two-plasmon threshold.
When the plasmon dispersion is positive (local-field-corrected results
at small $r_s$ and RPA) we expect small corrections, since 
the threshold behaviour is driven by the long-wavelength plasmon, whose
linewidth vanishes as $k^2$ for small $k$.
In the case of negative plasmon dispersion, the collective excitation with 
minimum energy has nonzero wavevector and therefore non-vanishing 
linewidth due to decay into multiple particle-hole pairs.
This indicats that the divergence in the two-pair spectrum found in the 
local-field-corrected results at large $r_s$ (see Figures
\ref{figconfim15} and \ref{figstlsrs}) is probably an artifact
and would be replaced by a smooth peak if plasmon damping were included.

\section{Summary}
\label{secconcl}
In this work we have given an extensive  discussion of the
exchange-correlation kernels $f_{xc}^{L,T}(\omega)$, both in 2 and in
3 spatial dimensions. We have presented an exact expression for the
kernels in terms of four-point response functions, and evaluated it
numerically within a non-perturbative approximate decoupling
scheme, which accounts for two-pair processes.
Our numerical results are qualitatively different from
previously known interpolations. In 3D we predict a threshold
behaviour, which can be understood as due to a simple
phase-space effect, i.e. the opening of the two-plasmon channel in the
two-pair spectrum. In 2D the same mechanism leads to a broad feature
in the spectrum.
We have also studied the influence of static-local-field corrections
on our results and found 
even more marked deviations from previous theories. 

We have obtained good
agreement with all known asymptotic behaviours, including the
newly-derived high-frequency limit of the transverse part and the
Monte Carlo results for the bulk modulus. 
Estimates for the shear modulus and viscosity have also been given.
Our results, which have been fitted to simple analytical formulas,
provide the entire 
input necessary for non-adiabatic TD-DFT computations.

\acknowledgments We wish to thank Prof.\ Giovanni Vignale, Dr.\ Carsten
Ullrich and Dr.\ Helga M.\ B\"ohm for very useful discussions, and
Dr.\ Lerwen Liu for 
sending us her data on $G(k)$ in STLS in the two-dimensional case. 
One of us (RN) acknowledges the award of a short-term
research grant by the FORUM/INFM Laboratory. 

\appendix

\section{Transverse response function for the non-interacting Fermi
  Gas in 2D and 3D}
\label{chizeroapp}
\label{applindhT}

The current-current response function for the non-interacting Fermi
gas is given by
\begin{equation}
\chi_{ij}^{0}({\bf k},\omega) = \sum_{{\bf q}}
{{\bf  q}^{i} {\bf q}^{j} \over m^2}
\frac {n_{{\bf q}+{\bf k}/2}^{F}-n_{{\bf q}-{\bf k}/2}^{F}}{\omega-
(\varepsilon_{{\bf q}-{\bf k}/2}-
\varepsilon_{{\bf q}+{\bf k}/2})+i\epsilon},
\end{equation}
where $n_{{\bf q}}^F=2\theta(k_F-|{\bf q}|)$
 is the Fermi distribution function and 
$\varepsilon_{{q}}=q^2/2m$ the free particle energy.
It is sufficient to evaluate the transverse component of
\begin{eqnarray}
  A_{ij}\left({\bf k},\omega\right) &=&
  -\pi\int\frac{d^{D}q}{\left(2\pi\right)^D} n^{F}_{q} 
  \frac{\left({\bf q}+\frac{1}{2}{\bf k}\right)^{i}}{m}
  \frac{\left({\bf q}+\frac{1}{2}{\bf k}\right)^{j}}{m}\nonumber\\
&&\times   \delta\left(\omega+
\varepsilon_{{\bf q}}-\varepsilon_{{\bf q}+{\bf k}}\right)
\end{eqnarray}
since ${\bf Im} \chi_{ij}^{0}({\bf k},\omega) = 
A_{ij}\left({\bf k},\omega\right)
- A_{ij}\left({\bf k},-\omega\right)$.  Angular integration leads to
\begin{equation}
A^T(k,\omega)=-\frac{1}{\pi m k}\int_{m|\omega-\varepsilon_k|/k}^{k_F}
d q\; \gamma_D\; q^D \; \theta(k_F-q) 
\end{equation}
where $\gamma_2=\sqrt{1-x^2}$, $\gamma_3=(1-x^2)/4$ and
$x=m|\omega-\varepsilon_k|/qk$.  It is convenient to work with the
reduced variables $z={k}/{2k_F}$ and $u={\omega m}/{k k_F}$.  After
integration we have
\begin{equation}
{\bf Im}\chi^0_T(k,\omega)=l_D\frac{n}{m z}\left(B_{+}-B_{-}\right)
\end{equation}
with $B_{\pm}=\theta\left(1-\mid u\pm z\mid\right) \left[1-\left(u\pm
    z\right)^{2}\right]^{\frac{D+1}{2}}$, $l_3=3\pi/32$ and $l_2=1/3$. 

The imaginary part of the ideal-gas transverse response function vanishes when
    $\left||\omega|-2kk_F/m\right|>\varepsilon_k$, as the longitudinal one.
 As a check of the 
present derivation we evaluate the third frequency moment sum rule,
which leads to 
\begin{equation}
\label{M3free}
M_{T}^0(k) = -{2\over\pi} \int_0^\infty \omega {\bf
  Im}\chi^0_T(k,\omega) d\omega = h_D 
{n\over m}\varepsilon_k \varepsilon_{{F}} 
\end{equation} 
with $h_3=4/5$ and $h_2=1$, in agreement with (\ref{M3Teq})
evaluated to zero order in the interaction potential $v_k^L$, i.e.
with $v_k^L=0$, ${\langle ke \rangle}={3\over
  5}\varepsilon_{F}$ in 3D and ${\langle ke \rangle}={1\over
  2}\varepsilon_{F}$ in 2D.

The real part is obtained from the imaginary one via the Kramers-Kronig
relation, which gives 
\begin{eqnarray}
{\bf Re}\chi^{0}_{T}(k,\omega)&=&
\frac{3n}{8 m z} 
\left[z^3+3u^2 z-{5\over 3}z\right.\nonumber\\
&& + \left.\frac{1}{4}E_+^{(3)}
-\frac{1}{4}E_-^{(3)}\right]
\end{eqnarray}
with $E^{(3)}_{\pm}=[(z\pm u)^2-1]^2\ln\left|\frac{u \pm z-1}{u \pm
    z+1}\right|$ in 3D, and
\begin{equation}
{\bf Re} \chi^0_{T}(k,\omega)=\frac{n}{3 m z}\left[2 z^{3}+
6 u^{2}z-3z-
E_{+}^{(2)}-E_{-}^{(2)}\right]
\end{equation}
with $E^{(2)}_\pm=[(z\pm u)^2-1]^{\frac{3}{2}}\theta(\mid z \pm u \mid-1)
{\mathrm{sgn}}(z\pm u)$ in 2D.

We examine now some limiting behaviours.  At long wavelength we get
\begin{equation} \chi^0_T(k\to0,\omega)= {M_T^0(k)\over\omega^2}
 + g_D {n\over m
  \omega^4}\varepsilon_k^2\varepsilon_{F}^2\,, \end{equation} 
where $g_3=48/35$, $g_2=1/2$ and $M_T^0(k)$ as defined in
  (\ref{M3free}).  At zero
frequency we have instead
\begin{eqnarray}
\chi^0_T(k,\omega=0) &=&
-{n\over m}\left[{5\over 8} -{3\over 32} {k^2\over k_F^2} \right.\nonumber\\
&&\left.
-{3\over 8} {k_F\over k}
\left({k^2\over 4 k_F^2}-1\right)^2 \ln\left|k -
2k_F\over k + 2k_F\right|
\right]  
\end{eqnarray}
in 3D, and 
\begin{eqnarray}
\chi_T^0(k,\omega=0)&=&-{n\over m}\left[1-{k^2\over 6 k_F^2} 
\right.\nonumber\\ 
&&+\left.{4 k_F\over 3k}
\left({k^2\over 4k_F^2}-1\right)^{3\over 2}\theta(k-2k_F)\right]  
\end{eqnarray}
in 2D. In both cases the $\omega\to 0, k\to 0$ limit depends on the order
in which the limits are taken, as in the longitudinal case.

\newbox\tmp\setbox\tmp=\hbox{{$k\;$}}
\section{Evaluation of the low-\box\tmp  expansion}
\label{appcommsmallk}

In this Appendix we evaluate to leading order in
$k$ the second term in the RHS of Eq.~(\ref{eqeqmotchiij}), which
obeys the equation of motion
\begin{eqnarray}
\omega^2{\bf Im}
\left\langle\!\left\langle {[{\bf j}_{{\bf k}},{H}]}; {[{\bf
        j}_{-{\bf k}},{H}]}
 \right\rangle\!\right\rangle_{\omega}
&=&-{\bf Im}
\left\langle\!\left\langle {[[{\bf j}_{{\bf k}},{{\rm K}+{\rm
          P}}],{{\rm K}+{\rm P}}]}; \right.\right.
\nonumber\\
&&
\hskip-1cm\left.\left.
{[[{\bf j}_{-{\bf k}},{{\rm K}+{\rm P}}],{{\rm K}+{\rm P}}]}
\right\rangle\!\right\rangle_{\omega},
\end{eqnarray}
where ${\rm K}$ and ${\rm P}$ denote the
kinetic and potential terms of the Hamiltonian
respectively. There are 16 terms in total, which correspond to the
different combinations of ${\rm K}$ and ${\rm P}$. Fortunately  
only 4 of them are relevant to our calculations, 
as is shown in the following.
Every commutator with ${\rm K}$ gives a factor of $k$, therefore terms 
containing more than two ${\rm K}$'s do not contribute to leading order in
$k$. The commutator $[[{\bf j}_{{\bf k}},{\rm P}],{\rm P}]$ vanishes, since
\begin{equation}
[{\bf j}_{{\bf k}},{{\rm P}}]=
[{\bf j}_{{\bf k}},\sum_{{\bf q}\ne0}v^L_{q}\rho_{{\bf q}}\rho_{-{\bf q}}]=
-\frac{1}{V}\sum_{{\bf q}\ne0}\frac{{\bf q}}{m}
\rho_{{\bf q}+{\bf k}}\rho_{-{\bf q}}
\end{equation}
and two density operators $\rho_{\bf q}$ commute.

The remaining terms can be written as
\begin{eqnarray}
&&\hskip-5mm -\frac{1}{\omega^{2}}
{\bf Im}\left\langle\!\left\langle
{[[{\bf j}_{{\bf k}},{{\rm K}}],
{{\rm P}}]+[[{\bf j}_{{\bf k}},{{\rm P}}],{{\rm K}}]}\right.\right.;
\nonumber\\
&&\hskip5mm \left.\left.{[[{\bf j}_{-{\bf k}},
{{\rm K}}],{{\rm P}}]+[[{\bf j}_{-{\bf k}},{{\rm P}}],{{\rm K}}]}
 \right\rangle\!\right\rangle_{\omega}\,.
\label{totaleq}
\end{eqnarray}
At this point we remark that the 
term ${\rm P}^{\mathrm{res}}=V^{-1}v^L_k\rho_{{\bf k}}\rho_{-{\bf k}}$ of the
Hamiltonian gives rise to a
singular contribution to the equation of motion for ${\bf j}_{\bf k}$,
\begin{equation}
\label{resterm}
[{\bf k}\cdot{\bf j}_{{\bf k}},{\rm P}^{\mathrm{res}}]=
\omega_{pl}^{2}\rho_{{\bf k}}- 
\frac{k^{2}}{mV}v^L_k \rho_{2{\bf k}}\rho_{-{\bf k}}.
\end{equation} 
In the first term in the RHS the factor $V^{-1}$ present
in the others has been compensated by the factor $\rho_0=N$. Therefore
this term has to be treated separately when
transforming the summations into integrals. We found it more
convenient to replace the Hamiltonian $H$ with  $\tilde{H}=H-{\rm
  P}^{\mathrm{res}}$ 
in the equation of motion for the response function, 
which allows to obtain the proper response
function $\tilde\chi$. This can be easily 
understood from
the diagrammatic definition of $\tilde\chi$ as the sum of all diagrams
that cannot be split by cutting a single interaction line: this
necessarily carries a factor $v^L_{{\bf q}={\bf k}}$,
which has been excluded from $\tilde{H}$. 
Using $\tilde H$ one also excludes all non-singular
contributions having a factor $v^L_{{\bf q}={\bf k}}$, as for example the
second term in 
the RHS of (\ref{resterm}). These terms however contain a factor
$V^{-1}$ and are irrelevant in the thermodynamic limit.

All considerations made previously hold with the new Hamiltonian, 
provided that we are
computing the proper current-current response in Eq.~(\ref{eqeqmotchiij}). The
commutators in (\ref{totaleq}) are given by
\begin{eqnarray}
{\left[{\left[{{\bf j}^{i}_{{\bf k}}},{{\rm K}}\right]},
{{\rm P}}\right]} &=&-\frac{1}{mV}\sum_{{\bf q}}
v^L_{q}\left(\frac{{\bf q}^{i}}{m}{\bf k}^{l}+
{\bf q}\cdot{\bf k}\delta_{il}\right)
{\bf j}^{l}_{{\bf q}+{\bf k}}\rho_{-{\bf q}} \,,
\nonumber\\
{\left[{\left[{{\bf j}^{i}_{{\bf k}}},
{{\rm P}}\right]},{{\rm K}}\right]} &=& \frac{1}{mV}
\sum_{{\bf q}} \left[v^L_{|{\bf q}+{\bf k}|}({\bf q}+{\bf k})^{i}-v^L_q
{\bf q}^{i}\right]\nonumber \\
&&\times({\bf q}+{\bf k}) \cdot {\bf j}_{{\bf q}+{\bf k}}\rho_{-{\bf q}}
\,.
\label{comms}
\end{eqnarray}
where terms containing a single $\rho_{{\bf k}}$ operator have been
neglected for reasons explained below.  From
Eqs.~(\ref{eqeqmotchiij}), (\ref{totaleq}) and (\ref{comms}) we get
\begin{eqnarray}
\label{final}
{\bf Im}{\tilde\chi}_{ij}({\bf k},\omega) &=&
\frac{1}{m^{2}V^{2}\omega^{4}}\sum_{{\bf q}{\bf q}'} 
{\bf Im}
\left\langle\!\left\langle {{\bf j}_{{\bf q}+{\bf k}}^{l}\rho_{-{\bf
          q}}}; 
{{\bf j}_{{\bf q}'-{\bf k}}^{l'}\rho_{-{\bf q}'}}
\right\rangle\!\right\rangle_{\omega}\nonumber\\
&\times& \Gamma^{il}({\bf q},{\bf k})\Gamma^{jl'}({\bf q}',-{\bf k})+o(k^{2})
\end{eqnarray}
where $\Gamma$ is given by
\begin{eqnarray}
\label{eqdefGamma}
\Gamma^{il}({\bf q},{\bf k}) &=& \left(v^L_{|{\bf q}+{\bf
k}|}-v^L_q\right){\bf q}^i
{\bf q}^l\nonumber\\
&+&v^L_q\left({\bf q}^l{\bf k}^i-{\bf k}^l{\bf q}^i -\delta_{il}
{\bf q}\cdot{\bf k}\right).
\end{eqnarray}
We remark that (\ref{final}) is valid  
for every Fourier-trasformable
interparticle potential, and is not restricted to $T=0$. 

The tensor $\Gamma$ is of first order in $k$ 
and the product of the two tensors
in (\ref{final}) is of second order, so that in the
response function we can safely set ${\bf k}=0$, leading to
Eq.~(\ref{eqesatta}). This also explains why terms 
containing one  $\rho_{{\bf k}}$ have been
neglected in the commutators (\ref{comms}): $\rho_{{\bf 0}}$ is
constant in time and its commutator with 
${\bf j}_{{\bf q}}\rho_{-{\bf q}}$ vanishes.

\section{Evaluation of the high-frequency limit via perturbative expansion}
\label{apppert}

Starting from Eq.~(\ref{eqesatta}) we evaluate the high frequency  
limit of $f_{xc}(\omega)$ to second order in perturbation theory.  
In this way we extend to the transverse case the results 
obtained by Glick and Long \cite{GlickLong} in 3D and by Holas and 
Singwi\cite{Holas89} in 2D  on the asymptotic behaviour of $f_{xc}$. 
The discussion  will also clarify the nature of the exchange processes which we
approximately included in the decoupling (\ref{eqimchil}) via the
factor $g_x$ of Eq. (\ref{eqdefgx}). 

The four-point response function can be written as
\begin{eqnarray}
\label{4point}
&&\hskip-1cm{\bf Im}
\left\langle\!\left\langle {{\bf j}_{{\bf q}}^{l}\rho_{-{\bf q}}}; 
{{\bf j}_{{\bf q}'}^{l'}\rho_{-{\bf q}'}}
\right\rangle\!\right\rangle_{\omega} =\nonumber\\ 
&=& -\pi \sum_{{\bf p}_{1}{\bf p}_{2}{\bf p}_{3}{\bf p}_{4}}
\frac{{\left({\bf p}_{1}+\frac{{\bf q}}{2}\right)^{l}}
\left( {\bf p}_{2}-\frac{{\bf q}'}{2} \right)^{l'}}{m^{2}}\nonumber\\
&&\times
\sum_{n}
\delta(\omega_{n0}-\omega)\langle{0}\vert
{c^{+}_{{\bf p}_{1}}c^{+}_{{\bf p}_{3}}
c_{{\bf p}_{1}+{\bf q}}c_{{\bf p}_{3}-{\bf q}}}\vert{n}\rangle\nonumber\\
&&\hskip1cm\times
\langle{n}\vert{c^{+}_{{\bf p}_{2}-{\bf q}'}c^{+}_{{\bf p}_{4}+{\bf q}'}
c_{{\bf p}_{2}}c_{{\bf p}_{4}}}\vert{0}\rangle
\end{eqnarray}
where $\vert0\rangle$ and $\vert n\rangle$ denote the exact
eigenstates of the system, and the spin index is implicit.  Since
each $\Gamma$ in (\ref{final}) contains a factor $v^L_q$, we can
evaluate the four-point response function at zero order in
perturbation theory, \emph{i.e.}  for the non-interacting electron
gas, so that $\omega_{n0}=({\bf q'}^2/m)+({\bf p}_4-{\bf p}_3)\cdot{\bf
  q}'/m$. 

The product of expectation values in the previous
equation is different from zero only when $p_1,p_2,p_3,p_4<k_F$, which
in the high-frequency limit implies $p_i \ll q \simeq q'
\simeq\sqrt{m\omega}$.  These considerations allow us to simplify
(\ref{4point}) as
\begin{eqnarray}
&&{\bf Im}
\left\langle\!\left\langle {{\bf j}_{{\bf q}}^{l}\rho_{-{\bf q}}};
{{\bf j}_{{\bf q}'}^{l'}\rho_{-{\bf q}'}}
\right\rangle\!\right\rangle_{\omega\to \infty}=\pi 
\frac{{\bf q}^l {\bf q}'^{l'}}{4 m^{2}}
\delta\left(
\frac{q'^{2}}{m}-\omega \right)\nonumber\\
&&\times \!\!\sum_{{\bf p}_{1}{\bf p}_{2}{\bf p}_{3}{\bf p}_{4}}\!\!
\langle{0}\vert
{c^{+}_{{\bf p}_{1}}c^{+}_{{\bf p}_{3}}
c_{{\bf p}_{1}+{\bf q}}c_{{\bf p}_{3}-{\bf q}}}
{c^{+}_{{\bf p}_{2}-{\bf q}'}c^{+}_{{\bf p}_{4}+{\bf q}'}
c_{{\bf p}_{2}}c_{{\bf p}_{4}}}\vert{0}\rangle
\end{eqnarray}
With $a,b,g,h<k_{F}$ and $c,d,e,f>k_{F}$, we have
\begin{eqnarray} 
\langle{0}\vert
{c^{+}_{a}c^{+}_{b}
c_{{c}}c_{{d}}}{c^{+}_{e}c^{+}_{f}
c_{g}c_{h}}\vert{0}\rangle&=&(\delta_{a,h}\delta_{b,g}
-\delta_{a,g}\delta_{b,h}) \nonumber\\
&&\times (\delta_{d,e} \delta_{c,f}-\delta_{c,e}\delta_{d,f}).
\end{eqnarray}
There are 4 terms overall, two of which are negative because of 
the anticommutation rules and mix together field operators belonging to
different density (or current density) operators. These terms, which
we call exchange terms, are neglected in the decoupling
(\ref{eqdecouplingabcd}) and 
carry an overall factor $-\frac{1}{2}$ due to spin (see below).
The first term $\delta_{a,h}\delta_{b,g}\delta_{d,e}\delta_{c,f}$ 
imposes the following constraints on the wave
vectors and spin indices,
\begin{eqnarray}
{\bf p}_{1}={\bf p}_{4} \;\; \sigma_{1}=\sigma_{4} &\;\;\;\;& 
{\bf p}_{1}+{\bf q}={\bf p}_{4}+{\bf q}' \;\; \sigma_{1}=\sigma_{4}
\,,\nonumber\\ 
{\bf p}_{2}={\bf p}_{3} \;\; \sigma_{2}=\sigma_{3} &&
{\bf p}_{3}-{\bf q}={\bf p}_{2}-{\bf q}' \;\; \sigma_{2}=\sigma_{3} \,.
\end{eqnarray}
The sum is done on ${\bf p}_1, \sigma_1$  ${\bf p}_2,\sigma_2$, and brings
a factor $N^2\delta_{{\bf q},{\bf q}'}$, with $N$ the number of particles. 
The second term $\delta_{a,g}\delta_{b,h}
\delta_{d,e} \delta_{c,f}$ gives
\begin{eqnarray}
{\bf p}_{1}={\bf p}_{3} \;\; \sigma_{1}=\sigma_{3}& \;\;\;\; & 
{\bf p}_{1}+{\bf q}={\bf p}_{4}+{\bf q}' \;\; \sigma_{2}=\sigma_{4}
\,,\nonumber\\ 
{\bf p}_{2}={\bf p}_{4} \;\; \sigma_{1}=\sigma_{4} &&
{\bf p}_{3}-{\bf q}={\bf p}_{2}-{\bf q}' \;\; \sigma_{2}=\sigma_{3}\,.
\end{eqnarray}
Unlike the first one, this second term fixes all the four $\sigma$'s, 
contributing with a factor $-\frac{N^2}{2}\delta_{{\bf q},{\bf q}'}$  which
is half the opposite of the first one. 

The remaining two terms are
obtained in an analogous way, and their overall contribution is
also $\frac{N^2}{2}\delta_{{\bf q},-{\bf q}'}$. 
Putting all terms together we get
\begin{eqnarray}
{\bf Im}
\left\langle\!\left\langle {{\bf j}_{{\bf q}}^{l}\rho_{-{\bf q}}};
{{\bf j}_{{\bf q}'}^{l'}\rho_{-{\bf q}'}}
\right\rangle\!\right\rangle_{\omega\to \infty} 
&=&\frac{\pi}{2}\frac{N^{2}}{m^{2}}\left(\delta_{{\bf q},{\bf q}'}+
\delta_{-{\bf q},{\bf q}'}\right)\nonumber\\
&&\hskip-.8cm \times\frac{{\bf q}^{l}{\bf q}'^{l'}}{4}
\delta\left(\frac{q^{2}}{m}
-\omega \right).
\end{eqnarray}
Substitution of the previous expression in Eq.~(\ref{eqesatta}) and small $k$
expansion lead to Eq.~(\ref{eqhighw}).

An estimate of higher--order corrections can be obtained from
additional applications of the equation of motion for ${\bf j}_{\bf q}$.
Third-order perturbation theory treats one of the two pairs to first
order and is irrelevant for large $q$ and $\omega$. To fourth order in
$v_k^L$, one obtains instead  three--pair contributions which contain
two additional 
commutators  with P in the last term of Eq.~(\ref{eqeqmotchiij}). 
Considering that the transferred momentum scales as $\sqrt{\omega}$ at
large $\omega$, these scale as $\omega^{-1-D}$ and therefore do not
affect the leading behaviour given by second-order perturbative expansion.


\begin{thebibliography}{10}
\bibitem[*]{presentaddress}
 Present address: Max-Planck-Institute for Mathematics in the Sciences,
  D-04103 Leipzig, Germany.

\bibitem{Fink}
A. vom Felde, J. Spr\"osser-Prou, and J. Fink, Phys. Rev. B {\bf 40},  10181
  (1989).

\bibitem{Larsen96}
B.~C. Larson, J.~Z. Tishler, E.~D. Isaacs, P. Zschack, A. Fleszar, and A.~G.
  Eguiluz, Phys. Rev. Lett. {\bf 77},  1346  (1996).

\bibitem{RungeGross}
E. Runge and E.~K.~U. Gross, Phys. Rev. Lett. {\bf 52},  997  (1984).

\bibitem{Gross96}
E.~K.~U. Gross, J.~F. Dobson, and M. Petersilka,  in {\em Density Functional
  Theory}, {\em Topics in Current Chemistry}, edited by R.~F. Nalewajski
  (Springer, Berlin, 1996).

\bibitem{Zangwill80}
A. Zangwill and P. Soven, Phys. Rev. Lett. {\bf 45},  204  (1980).

\bibitem{Zangwill81}
A. Zangwill and P. Soven, Phys. Rev. B {\bf 24},  4121  (1981).

\bibitem{Ekardt84}
W. Ekardt, Phys. Rev. Lett. {\bf 52},  1925  (1984).

\bibitem{Ekardt85}
W. Ekardt, Phys. Rev. B {\bf 31},  6360  (1985).

\bibitem{GrossKohn85}
E.~K.~U. Gross and W. Kohn, Phys. Rev. Lett. {\bf 55},  2850  (1985); erratum
  in Phys. Rev. Lett. {\bf 57}, 923 (1986).

\bibitem{IGK}
N. Iwamoto and E.~K.~U. Gross, Phys. Rev. B {\bf 35},  3003  (1987).

\bibitem{Dobson94}
J. Dobson, Phys. Rev. Lett. {\bf 73},  2244  (1994).

\bibitem{Dobson94b}
J. Dobson,  in {\em Density Functional Theory}, {\em NATO ASI}, edited by
  E.~K.~U. Gross and R.~M. Dreizler (Plenum, New York, 1994), p.\ 393.

\bibitem{Vignale95b}
G. Vignale, Phys. Rev. Lett. {\bf 74},  3233  (1995).

\bibitem{Vignale95a}
G. Vignale, Phys. Lett. A {\bf 209},  206  (1995).

\bibitem{Vignale96}
G. Vignale and W. Kohn, Phys. Rev. Lett. {\bf 77},  2037  (1996).

\bibitem{VignaleKohn96b}
G. Vignale and W. Kohn,  in {\em Electronic Density Functional Theory}, edited
  by J. Dobson, M.~P. Das, and G. Vignale (Plenum Press, New York, 1997).

\bibitem{VUC97}
G. Vignale, C.~A. Ullrich, and S. Conti, Phys. Rev. Lett. {\bf 79},  4878
  (1997).

\bibitem{UV98}
C.~A. Ullrich and G. Vignale, to be published  (1998).

\bibitem{Dabrowski86}
B. Dabrowski, Phys. Rev. B {\bf 34},  4989  (1986).

\bibitem{BCT96}
H.~M. B\"ohm, S. Conti, and M.~P. Tosi, J. Phys.: Condens. Matter {\bf 8},  781
   (1996).

\bibitem{Boi59}
D.~F. DuBois, Ann. Phys. (NY) {\bf 8},  24  (1959).

\bibitem{BoK69}
D.~F. DuBois and M.~G. Kivelson, Phys. Rev. B {\bf 186},  409  (1969).

\bibitem{NPS66}
B.~W. Ninham, C.~J. Powell, and N. Swanson, Phys. Rev. {\bf 145},  209  (1966).

\bibitem{HW}
M. Hasegawa and M. Watabe, J. Phys. Soc. Jpn. {\bf 27},  1393  (1969).

\bibitem{GlickLong}
A.~J. Glick and W.~F. Long, Phys. Rev. B {\bf 4},  3455  (1971).

\bibitem{Gas84}
W. Gasser, Zs. Phys. B {\bf 57},  15  (1984).

\bibitem{Gas92}
W. Gasser, Physica B {\bf 183},  217  (1992).

\bibitem{BMM91}
M.~E. Bachlechner, W. Macke, H.~M. Miesenb{\"o}ck, and A. Schinner, Physica B
  {\bf 168},  104  (1991).

\bibitem{BBS93b}
M.~E. Bachlechner, H.~M. B{\"o}hm, and A. Schinner, Phys. Lett. A {\bf 178},
  186  (1993).

\bibitem{Neilson}
D. Neilson, L. Swierkowski, A. Sj\"olander, and J. Szymanski, Phys. Rev. B {\bf
  44},  6291  (1991).

\bibitem{CNT97}
S. Conti, R. Nifos\`\i, and M.~P. Tosi, J. Phys.: Condens. Matter {\bf 9},
  L475  (1997).

\bibitem{NCT97}
R. Nifos\`\i, S. Conti, and M.~P. Tosi, Physica E {\bf 1},  188  (1998).

\bibitem{PinesNoz}
D. Pines and P. Nozi\`eres, {\em The Theory of Quantum Liquids} (Benjamin, New
  York, 1966), Vol.~1.

\bibitem{SingwiTosi}
K.~S. Singwi and M.~P. Tosi,  in {\em Solid State Physics}, edited by H.
  Ehrenreich, F. Seitz, and D. Turnbull (Academic, New York, 1981), Vol.~36,
  p.\ 177.

\bibitem{Lindhard54}
J. Lindhard, Mat.-Fys. Medd.-K. Dan. Vidensk. Selsk. {\bf 28},   No.~8  (1954).

\bibitem{Stern}
F. Stern, Phys. Rev. Lett. {\bf 18},  546  (1967).

\bibitem{Holas89}
A. Holas and K.~S. Singwi, Phys. Rev. B {\bf 40},  158  (1989).

\bibitem{footnotescreendpot}
 If the screened potential $\tilde{v}^L_k$ is considered, the
ratio  $\tilde v^L_{k_F}/\tilde v^L_q$ is still small at high density
(i.e. large $k_F$). At low density, on the other hand, the region
$q\ll k_F$ (i.e. $\omega\ll \varepsilon_F$) shrinks to zero. 

\bibitem{Landau6}
L.~D. Landau and E. Lifshitz, {\em Mechanics of Fluids}, Vol.~6 of {\em Course
  of Theoretical Physics}, 2nd ed. (Pergamon Press, Oxford, 1987).

\bibitem{CeperleyAlder}
D.~M. Ceperley and B.~J. Alder, Phys. Rev. Lett. {\bf 45},  566  (1980).

\bibitem{VWN80}
S.~H. Vosko, L. Wilk, and M. Nusair, Can. J. Phys. {\bf 58},  1200  (1980).

\bibitem{Rapisarda96}
F. Rapisarda and G. Senatore, Aust. J. Phys. {\bf 49},  161  (1996).

\bibitem{Landau7}
L.~D. Landau and E. Lifshitz, {\em Theory of Elasticity}, Vol.~7 of {\em Course
  of theoretical Physics}, 3rd ed. (Pergamon Press, Oxford, 1986).

\bibitem{fxclungobetabar}
If complete consistency with $K_{xc}^{\mathrm{MC}}$ is desired, one simple
  possibility would be to multiply ${\bf Im} f_{xc}^{L,T}(\omega)$ by a
  constant $\beta$, chosen so as to satisfy $K_{xc} = K_{xc}^{\mathrm{MC}}$,
  i.e. $$\beta= {K_{xc}^{\mathrm{MC}} - n^2\left[ f_{xc}^L(\infty) - 2
  {D-1\over D} f_{xc}^T(\infty)\right] \over K_{xc} - n^2\left[
  f_{xc}^L(\infty) - 2 {D-1\over D} f_{xc}^T(\infty)\right]}\,. $$

\bibitem{STLS}
K.~S. Singwi, M.~P. Tosi, R.~H. Land, and A. Sj\"olander, Phys. Rev. {\bf 176},
   589  (1968).

\bibitem{VS}
P. Vashishta and K.~S. Singwi, Phys. Rev. B {\bf 6},  875  (1972).

\bibitem{Jonson76}
M. Jonson, J. Phys. C {\bf 9},  3055  (1976).


\end{thebibliography}

\begin{figure}
\centerline{\psfig{width=0.98\columnwidth,figure=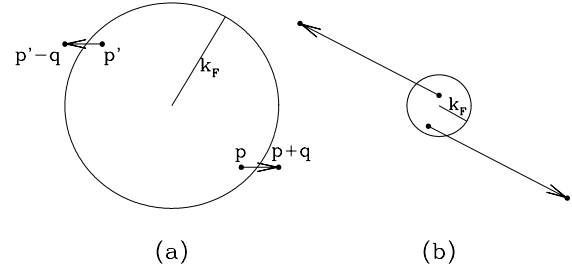}}
\caption{Structure of two-pair excitations in a Fermi liquid for
  energies much smaller (a) and much bigger (b) than the Fermi
  energy. The arrows join the electron and hole of each pair, their
  length is the transferred wavevector ${\bf q}$.} 
\label{figfs}
\end{figure}

\begin{figure}
\centerline{\psfig{width=0.98\columnwidth,figure=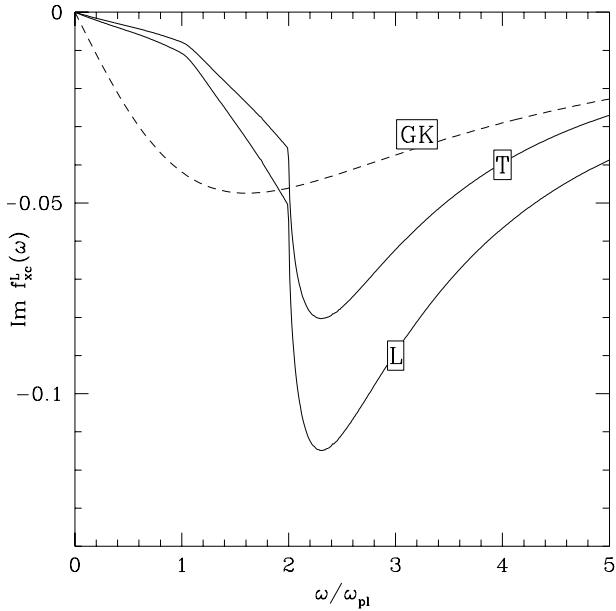}}
\caption{Imaginary part of $f^{L,T}_{xc}(\omega)$ in 3D at $r_s=3$ in
units of $2{\omega_{pl}}/n$, as functions of
$\omega/{\omega_{pl}}$. The dashed line gives 
the Gross-Kohn \protect\cite{GrossKohn85,IGK} interpolation scheme.}
\label{figim3}
\end{figure}

\begin{figure}
\centerline{\psfig{width=0.98\columnwidth,figure=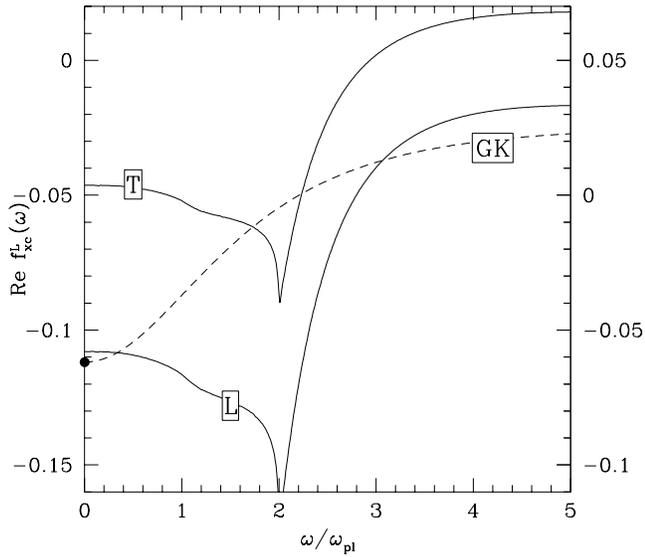}}
\caption{Real part of $f^{L,T}_{xc}(\omega)$ in 3D at $r_s=3$. Notations and
units are as in Fig.~\protect\ref{figim3}. The dot on the left axis marks
$K_{xc}^{\mathrm{MC}}/n^2$. The scale for the transverse component is
on the right-hand axis.}
\label{figre3}
\end{figure}

\begin{figure}
\centerline{\psfig{width=0.98\columnwidth,figure=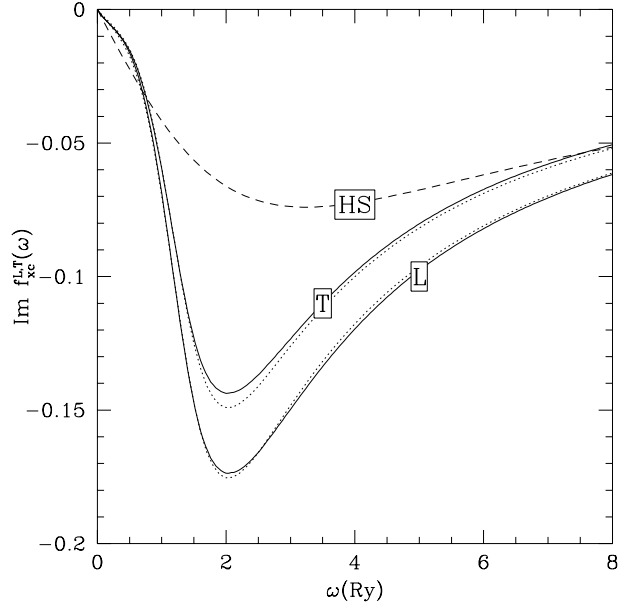}}
\caption{Imaginary part of $f^{L,T}_{xc}(\omega)$ in 2D at $r_s=3$ in
units of $Ry/n$, as functions of $\omega$ (in $Ry$). The dashed line gives
the Holas-Singwi \protect\cite{Holas89} interpolation and the dotted
lines are the fit discussed in the text.}
\label{fig2dim2}
\end{figure}

\begin{figure}
\centerline{\psfig{width=0.98\columnwidth,figure=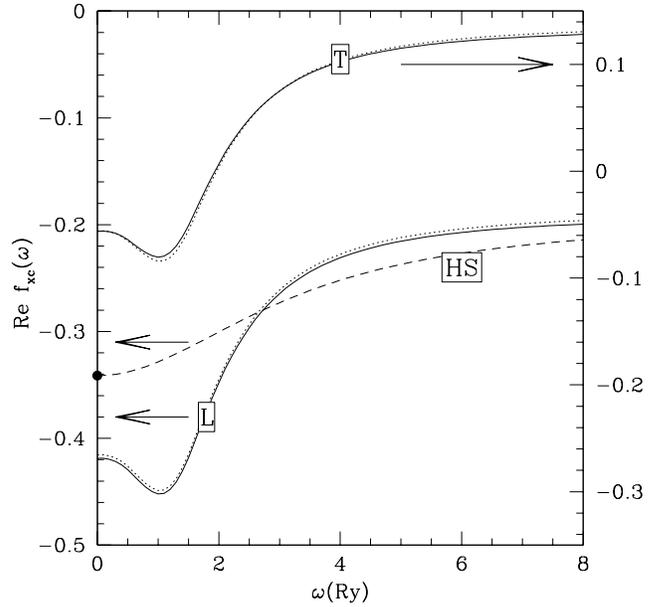}}
\caption{Real part of $f^{L,T}_{xc}(\omega)$ in 2D at $r_s=3$ in
units of $Ry/n$, as functions of $\omega$ (in $Ry$), the scale for the
transverse component being on the right-hand axis. The dashed line gives
the Holas-Singwi \protect\cite{Holas89} interpolation and the dotted
lines show the fit discussed in the text. The dot on the left axis marks
$K_{xc}^{\mathrm{MC}}/n^2$.}
\label{fig2dre2}
\end{figure}

\begin{figure}
\centerline{\psfig{width=0.98\columnwidth,figure=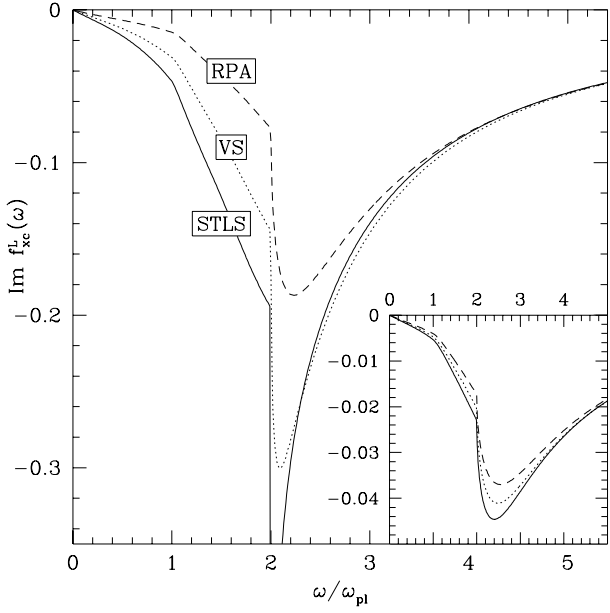}}
\caption{Imaginary part of $f^L_{xc}(\omega)$ in 3D in
units of $2{\omega_{pl}}/n$, as a function of $\omega/{\omega_{pl}}$  
at $r_s=5$ (main figure) and $r_s=1$ (inset). We plot results obtained
using RPA response functions (dashed 
curve), STLS response functions (full curve) and VS response
functions (dotted curve).}
\label{figconfim15}
\end{figure}

\begin{figure}
\centerline{\psfig{width=0.98\columnwidth,figure=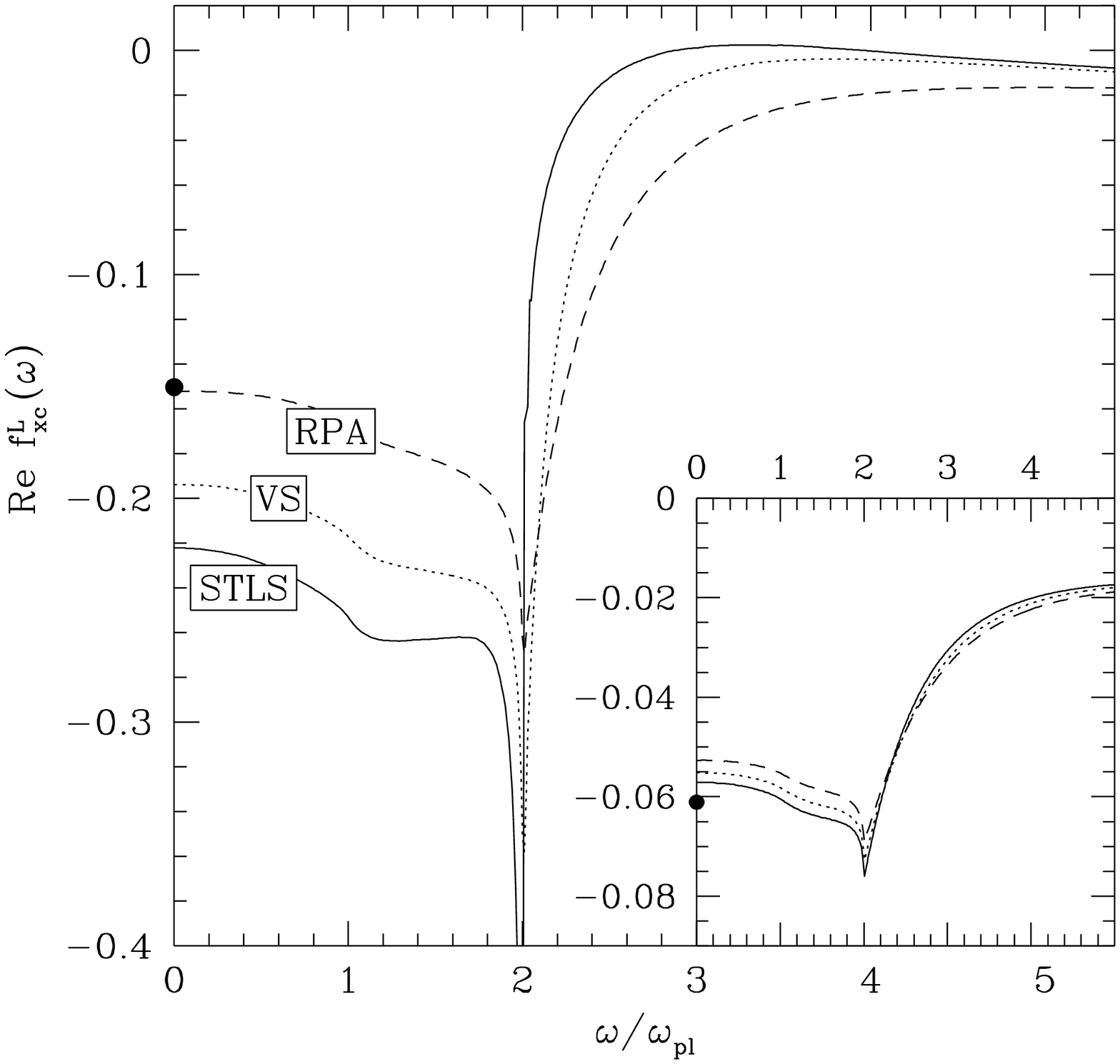}}
\caption{Real part of $f^L_{xc}(\omega)$ in 3D in
units of $2{\omega_{pl}}/n$, as a function of $\omega/{\omega_{pl}}$
at $r_s=5$ (main figure) and $r_s=1$ (inset).
We plot results obtained using RPA response functions (dashed
curve), STLS response functions (full curve) and VS response
functions (dotted curve). The dots mark 
$K_{xc}^{\mathrm{MC}}/n^2$.}
\label{figconfre15}
\end{figure}

\begin{figure}
\centerline{%
\psfig{figure=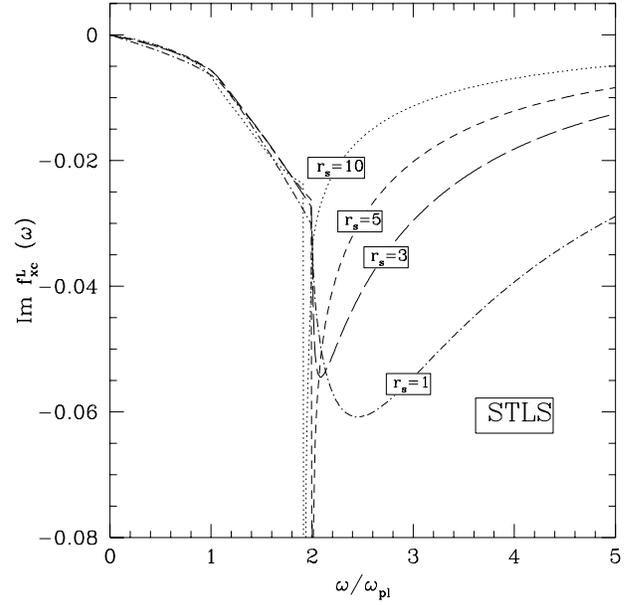,width=0.98\linewidth}}
\caption{Imaginary part of ${f_{xc}^L}(\omega)$ in 3D, in units of
$2r_s^{-3/2}{\omega_{pl}}/n$, computed with STLS response functions, at
various values of $r_s$.} 
\label{figstlsrs}
\end{figure}

\begin{figure}
\centerline{\psfig{figure=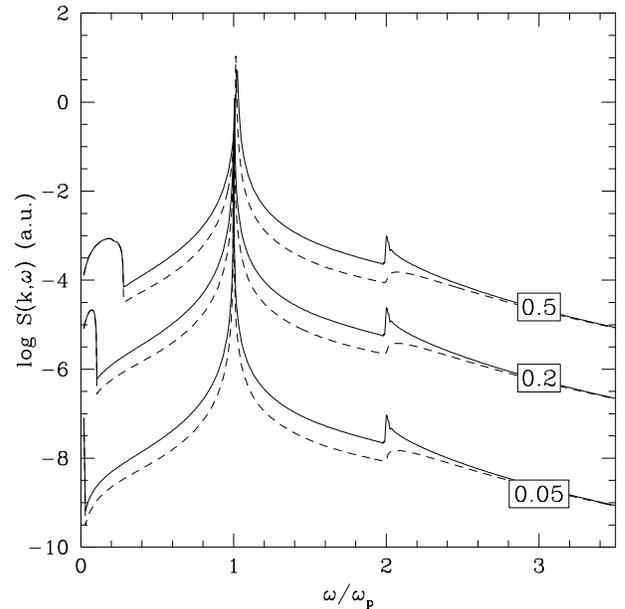,width=0.98\linewidth}}
\caption{Dynamic structure factor $S(k,\omega)$ at $r_s=5$ as a function
of $\omega/{\omega_{pl}}$ shown in the flags on a semilogarithmic
scale
 at various values of $k r_s a_B$, as obtained from STLS (full curves)
and  RPA (dashed curves)
calculations.} 
\label{figskwstlsrpa}
\end{figure}

\begin{figure}
\centerline{\psfig{width=0.98\columnwidth,figure=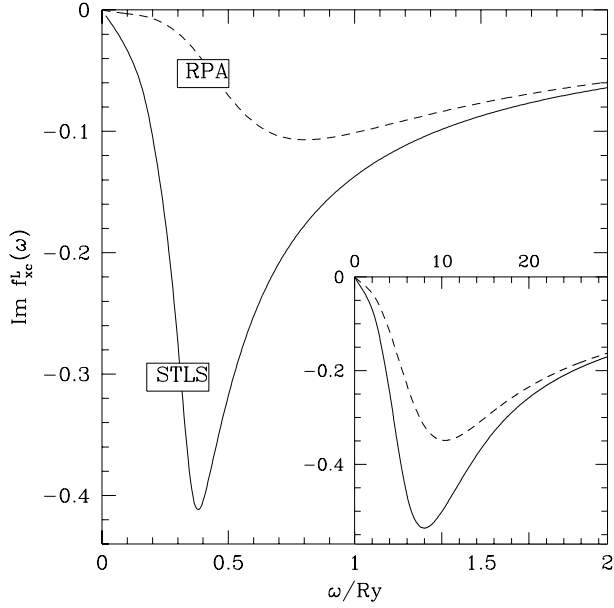}}
\caption{Imaginary part of ${f_{xc}^L}(\omega)$ in 2D in
units of $Ry/n$, as a function of $\omega$ (in $Ry$)
at $r_s=6$ (main figure) and $r_s=1$ (inset).
We plot results obtained using RPA response functions (dashed
curves) and STLS response functions (full curves).}
\label{fig_rpastls2dim}
\end{figure}

\begin{figure}
\centerline{\psfig{width=0.98\columnwidth,figure=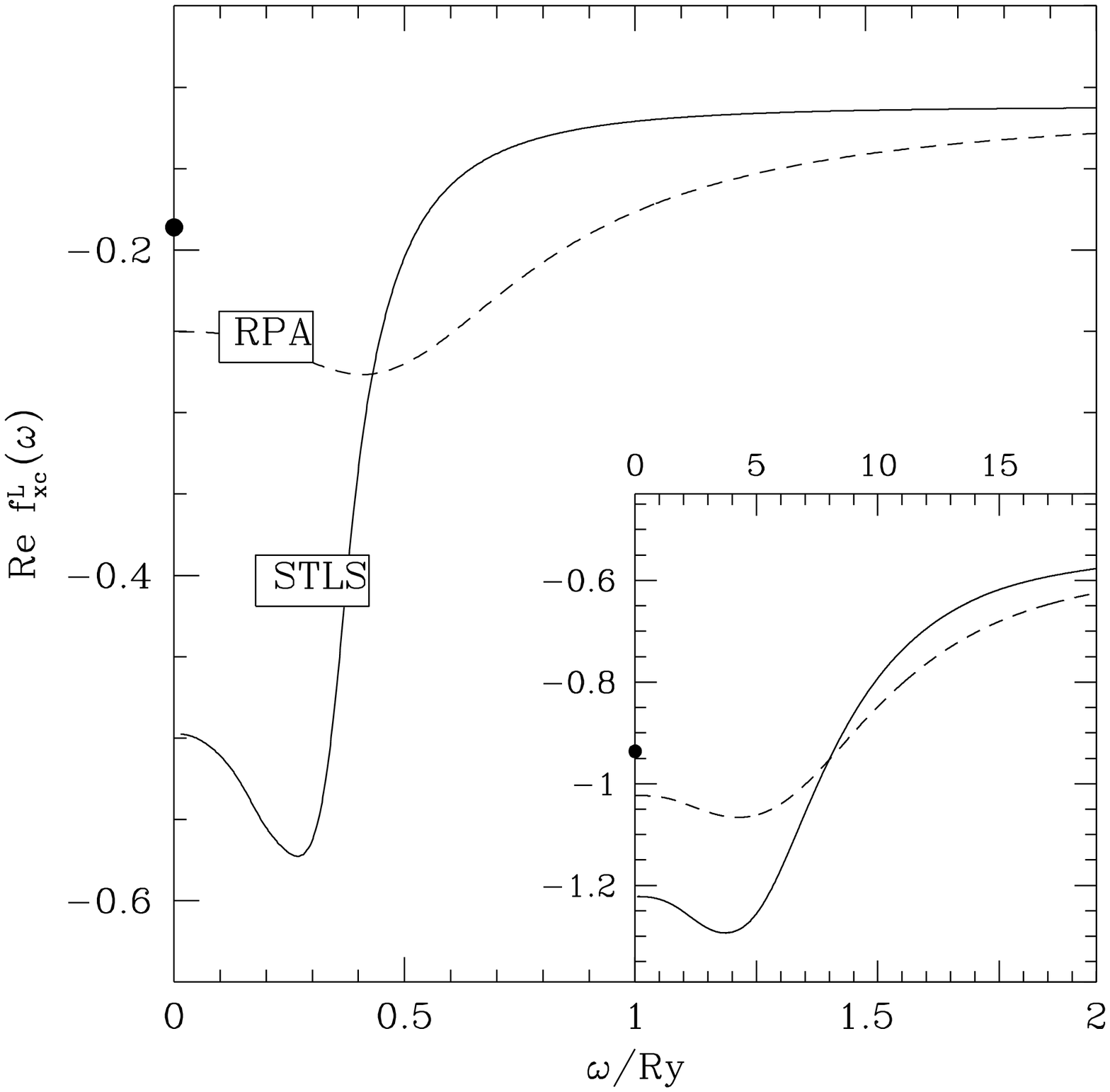}}
\caption{Real part of ${f_{xc}^L}(\omega)$ in 2D in
units of $Ry/n$, as a function of $\omega$ (in $Ry$)
at $r_s=6$ (main figure) and $r_s=1$ (inset).
We plot results obtained using RPA response functions (dashed
curve) and STLS response functions (full curves). The dots
mark $K_{xc}^{\mathrm{MC}}/n^2$.}
\label{fig_rpastls2dre}
\end{figure}

\clearpage
\mediumtext

\begin{table}
\caption{Exact high-$\omega$ limits of $f_{xc}^{L,T}(\omega)$ in 3D 
and bulk modulus $K_{xc}^{\mathrm{MC}}$ obtained from the
Monte Carlo equation of state; shear viscosity
$\eta$ and shear and bulk moduli ($\mu_{xc}$ and
$K_{xc}$) obtained from the RPA treatment of two pair processes. 
$f_{xc}$ is in units of
 $2\omega_{pl}/n$, $\eta$ in units of $n$, $K_{xc}$ and $\mu_{xc}$ in units
 of $2\omega_{pl} n$. Values in atomic units ($a_B=m=e^2=1$) can be
obtained from 
$\eta^{AU} =  3/(4 \pi r_s^3)\eta^{\mathrm{tab}}$,
$(K,\mu)^{AU} =  3^{3/2} r_s^{-9/2}/(2\pi) (K,\mu)^{\mathrm{tab}}$,
where $^{\mathrm{tab}}$ denotes the tabulated values.
The last reported decimal figure is likely
 to be affected by numerical inaccuracies.}
\label{tabvisco3D}
\label{tablefxcasy}
\begin{center}
\begin{tabular}{dddddddd}
$r_s$ & $K_{xc}^{\mathrm{MC}}$ &  $f_{xc}^L(\infty)$ & 
$f_{xc}^T(\infty)$ & $\eta$  &
$\mu_{xc}$ & $K_{xc}$ \\ \hline 
0.5& -0.04246 & -0.01794 & 0.0177 & 0.0029 & 0.0065 & -0.0425\\ 
1  & -0.0611 & -0.0216 & 0.0284  & 0.0062  & 0.0064 & -0.0612\\ 
2  & -0.0891 & -0.0252 & 0.0457  & 0.012  &  0.0052 & -0.0896\\ 
3  & -0.1119 & -0.0280 & 0.0600  & 0.017  & 0.0037  & -0.1128\\ 
4  & -0.1320 & -0.0308 & 0.0724  & 0.021  & 0.0020  & -0.1335\\ 
5  & -0.1503 & -0.0338 & 0.0835  & 0.024  & 0.0002  & -0.1525\\ 
6  & -0.1674 & -0.0370 & 0.0935  & 0.027  & -0.0018 & -0.1702\\ 
10  & -0.2276 & -0.0518 & 0.1267 & 0.034  & -0.010  & -0.233\\  
15  & -0.2917 & -0.0725 & 0.1587 & 0.040  & -0.023  & -0.301\\ 
20  & -0.3483 & -0.0939 & 0.1847 & 0.044  & -0.036  & -0.361\\  
\end{tabular}
\end{center}
\end{table}

\begin{table}
  \caption{Exact high-$\omega$ limits of $f_{xc}^{L,T}(k,\omega)$ in 2D
and bulk modulus $K_{xc}^{\mathrm{MC}}$ obtained from the
Monte Carlo equation of state; shear viscosity
$\eta$ and shear and bulk moduli ($\mu_{xc}$ and
$K_{xc}$) obtained from the RPA treatment of two-pair processes. 
$f_{xc}$ is in units of
 $Ry/n$, $\eta$ in units of $n$, $K_{xc}$ and $\mu_{xc}$ in units
 of $Ry\cdot n$. 
Values in atomic units can be obtained from
$\eta^{AU} = 1/( \pi r_s^2) \eta^{\mathrm{tab}}$,
$(K,\mu)^{AU} = 1/(2\pi r_s^2) (K,\mu)^{\mathrm{tab}}$,
where $^{\mathrm{tab}}$ denotes the tabulated values. 
The last reported decimal figure is likely
 to be affected by numerical inaccuracies.} 
\label{tabvisco2d}
\begin{center}
\begin{tabular}{ddddddd}
$r_s$ & $K_{xc}^{\mathrm{MC}}$ & ${f_{xc}^L}(\infty)$ &
${f_{xc}^T}(\infty)$ & $\eta$ &
$\mu_{xc}$ & $K_{xc}$ \\ \hline 
1  & -0.9360 & -0.5499 & 0.3372 & 0.018 & -0.064& -0.959
 \\
2  & -0.4912 & -0.2750 & 0.1916 & 0.029 & -0.064& -0.514
 \\
3  & -0.3413 & -0.1933 & 0.1330 & 0.035 & -0.058& -0.363
 \\
4  & -0.2649 & -0.1535 & 0.1010 & 0.040 & -0.054& -0.285
 \\
5  & -0.2180 & -0.1294 & 0.0810 & 0.043 & -0.050& -0.236
 \\
6  & -0.1860 & -0.1128 & 0.067 & 0.045 & -0.047& -0.203
 \\
10  & -0.1191 & -0.0768 & 0.0395 & 0.050 & -0.039& -0.132
 \\
15  & -0.0833 & -0.0560 & 0.0257 & 0.052 & -0.033& -0.094
 \\
20  & -0.0645 & -0.0445 & 0.0189 & 0.054 & -0.028& -0.073
 \\
\end{tabular}
\end{center}
\end{table}

\narrowtext
\begin{table}
  \caption{
Interpolation parameters according to Eq.~(\protect\ref{eqfxcfit2d}),
in 2D. The last reported decimal figure is likely
 to be affected by numerical inaccuracies.} 
\label{tabparfit2d}
\begin{center}
\begin{tabular}{dddddd}
$r_s$ & $ 10^{-3} c_0 r_s^{5/2} $ & 
$ c_1 r_s^2$ & $c_2$ & $c_3$ &
$c_4$  \\ \hline 
1  
 & 62.7&1.10   & 9.94 &37.4 & 6.84\\
2  
  & 2.90& 59.2& -1.74& 4.62& 7.70\\
3  
 & 1.73& 34.9& -3.79& 12.5& 15.1\\
4  
 & 1.44& 28.6& -3.45& 15.8& 30.3\\
5  
 & 1.18& 22.5& -2.64& 16.3& 47.9\\
6  
 & 0.943& 17.3& -2.04& 15.8& 66.2\\
10  
  & 0.458& 7.23& -1.0& 13.1& 150\\
15  
 & 0.250& 3.42& -0.524& 10.9& 276\\
20  
 & 0.160& 1.96& -0.313& 9.52& 430\\
\end{tabular}
\end{center}
\end{table}

\end{document}